# Reengineering multi tiered enterprise business applications for performance enhancement and reciprocal or rectangular hyperbolic relation of variation of data transportation time with row pre-fetch size of relational database drivers

[1]Sridhar Sowmiyanarayanan

[1]Technology Excellence Group, Banking and Financial Services Technologies, Tata Consultancy Service Ltd, Bangalore, Karnataka, India

**Abstract**

In a traditional multitier applications performance bottlenecks can be in user interactions level or network latency or data access or business logic level. The solutions as changes or tuning parameters can be applied at architect, design, framework or algorithm or at coding level. This paper highlights an inquisitive, experimental, top down, tear apart, drill down and analytical approach across two aspects one across end to end process flow and data flow on those specific use cases or scenarios requiring performance improvement and another across layers of abstraction like architecture, framework, design, logic and coding. Re engineering for performance gain requires identifying hot spots on both aspects viz which architectural, design decision or which processing or data flow stage is having performance issue. Once identified one can further drill down and identify root cause and also can find solution as a change. To help the application owner in decision making process, the analysis outcome should have tuning parameters, relationship between them, optimum values, tradeoffs on each changes, effort, risks, cost and benefits for incorporating each change. Following the above mentioned approach on a system in production with large enterprise we could drill down to a rectangular hyperbolic or reciprocal relation between elapsed time to transport all records retrieved from a query and the number of records being pre fetched (pre fetch size) and cached in the data base client application by the database driver in each trip. Because of the reciprocal nature , we could observe that when the pre fetch size is low drastic reduction in elapsed time could be obtained even for a small increase in pre fetch size, whereas when the pre fetch size is high the gain in performance is not so significantly high even for larger increase in pre fetch size.

## Introduction

There is growing need and challenges to re engineer existing business applications in production to improve on quality attributes like performance and scalability. Business applications systems built without focusing on non functional requirements or quality attributes and their future growth in demand are facing the need and challenges of reengineering for improving quality attributes. Some quality attributes like reliability requirement may not change or increase over a period of time but performance and scalability demand may increase due to increasing number of concurrent users of the system or increasing data volume which needs to be processed by the system. Among a large set of different business applications though there may be many reasons for performance bottlenecks in traditional n tier system, the most common performance bottleneck area could be in data access layer or avoidable high memory footprint of the application.

A general approach to re engineer for performance improvement is to elicit and extract the implemented architect and design. Understand the various stages of process and data flow focusing on those scenarios or use cases for which performance gain is required. One can time profile across various processing and data flow stages to identify hot spots and can further drill down in to details with experimentation and measurements to identify root cause. Once root cause is known solutions can be found as changes. Changes can be at various levels of abstractions like at architecture, design, logic etc. Estimates of performance gain, effort required for incorporating each change, tradeoffs for each changes will help the application owner to decide on cost benefit and other impacts to incorporate each recommended changes. Such analytical or experimental relations between tuning parameters will be greatly helpful because such relations need not be specific to particular application but can be generic and same analytical relations can be reused for tuning similar applications with similar environment in which such relations are valid.

Re engineering for performance gain and quantitative analysis both experimental and analytical on data transportation between application server and database done on a java j2ee banking application for bank's customers to view online reports related to their assets under banks custody is detailed here as a case study application. The outcome of the case study as a reciprocal relationship between total time (T) to transport query results of N records and the pre fetch size (f) is discussed. Though the relation between T and f may have other factors, we could approximate to the reciprocal relation and rectangular hyperbolic trend as a dominant factor.

## 1. Need for performance enhancements

Many enterprises had over the years built software applications to meet their operational, transactional or for customer services. Over the period of time, the load in terms of number of users or size of data being processed had grown but the built system had not scaled proportionately. This resulted in low performance of the application and or its inability to scale to meet the increasing demand. Some of the reasons any enterprise in general or banking and financial sector to opt for performance or scalability enhancements of their applications are listed below.

1. The technical and application architecture would not have been planned appropriately taking in to account the non functional requirements properly as the development team would have focused primarily on functional requirements. Other common reason could be that the system might have evolved over few years in ad hoc manner rather than planned, architected, well designed and built.

2. The load in terms of number of users or number of files or number data records to process would have increased recently but the system could not meet new increased load.

3. The load in terms of size of file and or size of messages to process would have increased recently.

4. Additional functionalities were or needed to be rolled out but the system could not perform or scale to the additional functional needs.

## 2. Common Non Functional Requirements which can change with time with higher demand

For many business applications, the demand for some quality attributes or non functional requirements like reliability may not change over a period of time whereas demand for other quality attributes like performance may increase with time. The three most common qualities of service, the demand for which can increase over a period of time with some illustrative examples are

### 2.1 Performance

    a. User response time in the case of interactive applications or

    b. Processing time in case of batch (non interactive) applications.

### 2.2 Scalability

2.1.1. Number of online concurrent users to support in the case of interactive applications.

2.1.2. Number of records or files or messages to process in the case of batch applications.

2.1.3. Size of file or size of message being processed.

### 2.2. Throughput

For example number of work items completed over period of time. Work items for example can be number of trades to be prized or number of trades to be settled or number of transactions to be completed or number of pairs of records to be reconciled.

## 3. Possible performance problem areas

In traditional N tier client application server database server enterprise business applications, performance bottlenecks can be in any or more of

3.1. Inefficient data access from database

3.2. Inefficient data transport between client and application server or application server and database

3.3. Inefficient I/O operations like file I/Os

3.4. Network latency

3.5. Inefficient application logic and inefficient algorithms

3.6. Resource contention like contention for CPU and memory as clients or user sessions, server executables or threads waiting for their turn for these resources

3.7. Higher memory foot print or frequent memory page faults resulting in higher percentage of disk based I/O.

Above are not comprehensive but common performance bottlenecks. Among these inefficient data access and inefficient data transport from database are the most common cause of performance bottlenecks in many applications.

## 4. General approaches to reengineer for performance optimization.

Assume that a banking enterprise engages a software services vendor to enhance the performance of its existing web client-application server-data base tiered application in production. Though there can be multiple ways and approaches any vendor can adopt, following describes one possible approach and steps to reengineer the application to improve on performance.

4.1. Elicit and enumerate the list of scenarios and use cases which are having low quality of services such as low performance and get a scope of problem areas.

4.2. Elicit and understand the implemented architecture, high level design, high level data and process flow and get a high level bird's eye view of functional, technical, process and data flow overview.

4.3. Breakup the high level flow in to various smaller processing stages. Instrument the code or use profilers and tools and do test run to record elapsed time breakups across various stages of the application flow.

4.4. Identify where maximum time is spent across various stages.

4.5. Identify list of root causes for the performance bottlenecks within identified low performing stages

4.6. Find solution as changes to rectify the root causes.

4.7. Changes can range from simple parameter changes to changes at code, logic design and at architecture level.

4.8. Measure or estimate benefits for each parameter changes and get a quantitative and analytical detail on each parameter.

4.9. Parameters don't impact the performance in isolations and changes in one parameter may influence the benefit derived due to changes made in other parameters.

4.10. For each change identified as a solution to improve performance, estimate quantitatively or qualitatively the following

4.10.1. Tradeoffs

4.10.2. Cost of applying the changes and benefits if the changes are incorporated

4.10.3. Risks associated with each change.

4.10.4. For example using memory caching to store all application data instead of persistent database can be a suggested change to improve performance, but the trade off are 4.10.4.1. Between performance and scalability to high data volume.

4.10.4.2. Licensing cost of the caching product and cost for the effort to change and test.

4.10.4.3. Risk of reliability in terms of data loss if there are system crashes/failures and if there are no failover recovery mechanism implemented for data in volatile memory.

## 5. Various levels of performance problems and solutions

The problem and solution can be at various levels of abstraction as given below.

*Table 1: Illustrative performance problems at various levels and sample solution of a typical multi tiered application*

| Levels of abstraction | Example problem areas | Example change scenarios to gain performance |
|---|---|---|
| Architecture level | Disk based I/O | Caching can be used to reduce latency. |
| | Application access data from remote locations | Deployment architecture can be changed to co-locate application and data in same geography to avoid networked data access. |
| | User responses found to slow down, when large number of simultaneous users made request or processing time slows down when large number of data records has to be processed. | Request processing can be parallelized by more server instances and load balancing. If processing of one data record is independent of other data record, then data records processing can be parallelized by many server instances. |
| Design level | I/O and CPU are sequential though there is opportunity to do both concurrently. | Threads can be used to concurrently do I/O and processing |
| Framework level | Framework has too many layers: Assume that the used framework introduced too many layers of indirection and data conversion and transportation. For example used framework marshals and un marshal's data across network, converts data from flat file to XML to java object to data base objects using ORM (object relational mapping software). | More appropriate framework can be chosen to avoid or minimize data marshalling, un marshalling, conversions and transportations. |
| Logic level | Used algorithm is inefficient | Algorithm can be changed to be more efficient and optimal. |
| Coding level | Loop invariants: Reading end of day currency exchange rate in a million trade record loop to convert price of each trade from Euro to US$. | Onetime currency exchange rate can be read outside the trade record loop and the read value can be used inside the loop, i.e. loop invariant statements can be taken outside loop to avoid repeated execution. |
| | Resource contention: For example process or threads waiting more than required duration for a shared resource like data base connectivity. Another example is inefficient resource locking and release mechanism among multiple process or threads. | Resource management can be optimized: Resources like database connection or locks can be released immediately after use. Resource pools like connection pool or thread pools can be used to minimize time on resource re recreation every time. |
| Resource/Infrastructure consumption/utilization level | Un optimal resource utilization: Assume that the system has 2 CPUs or the system is a dual core system and there are 2 stages of data processing in | The strategy can be changed to spawn 2 instances of process A to run in parallel to complete stage 1 first and then to spawn 2 instances of process B to run |

|  | an application and the application spawned 2 processes A and B spawned to run a process in each of the CPU. If the total time is 3 hours and if 1$^{st}$ stage is processed by process A and is finished in first 1 hour, then remaining 2 hours only one among 2 CPUs are being consumed wasting 1 CPU resource for 2 hours. | stage 2, thereby consuming both the CPU resource during the entire duration of execution to maximize resource utilization effectively. |
|---|---|---|
|  | High memory footprint: Assume that the designed system's memory foot print of each user/session is avoidably high and because of accumulating memory when 50+ users logged in, the application slows down due to very high memory usage. | Memory footprint can be reduced to as minimum as possible for each user/session so that the side effects of high memory consumption slowing down system performance can be minimized. |

## 6. Quantitative estimation and impact study of performance gain from multiple causes and respective solutions

There can be multiple causes and multiple solutions for each cause to the given performance problem. Many applications in general may have performance bottlenecks due to various reasons at various levels and stages and performance gain is possible from respective solutions. Thus the opportunity to make changes and gain performance may be scattered at multiple points from beginning of processing to end of processing throughout the application. For example following changes specific to a specific application may yield performance benefits. Co locating database and application server, few core logic changes, data writes changes to batch writes and data read changes to reduce number of round trips between application server and database server.

Each of the above changes may bring some performance benefits. It is essential to analyze, model, measure or estimate quantitatively the amount of performance gain each change may yield, the tradeoffs with each changes, the cost and effort of each changes, relative benefits, risks of introducing bugs and functional and operational impacts has to be studied. According to the analysis and impact study outcome, implementation of proposed changes can be undertaken. Thus the recommendation should not only contain solutions or tuning parameters or changes and benefits but also should provide adequate support data like quantitative benefits (performance gain) achievable by each change, tradeoffs, effort, cost of making those changes, complexity, risks for each change so that customer get adequate information from the recommendations to do cost benefit analysis and make an informed and calculated decision.

## 7. Management issues and cost benefit analysis

Most of the time performance issues are identified at much later stage of developments like

  7.1. During load and volume testing

  7.2. Unexpected peak demand during production

Thus performance engineers are left with limited options of

  7.3. Limited timelines to fix the performance issue

  7.4. Constrained to make only minor architectural or design changes or deviations as larger changes may require

  7.4.1. Longer testing cycles

  7.4.2. Higher risk of introducing regression bugs.

  7.4.3. Higher effort and hence requires more time for the change

  7.4.4. Longer effort and hence higher cost of change

  7.4.5. Larger change to the existing system implies low realization of return on investments made on the existing system.

 7.5. Engineers can make only limited technical changes

  7.5.1. Which requires less effort in terms of time and manpower

  7.5.2. Less risk in terms of breaking functionalities or causing new functional bug.

  7.5.3. Lower cost by avoiding new third party products, licenses or IPs

## 8. Case study illustrating approach, drill down to root causes, change parameters and relationship between parameters.

Below sections illustrates with a case study application, how a top down approach described above has identified data access as a cause. How a further breakup of data access in to smaller stages within data access pointed to data transportation and how a further drill down in to data transportation pointed to low pre fetch size (record numbers) as one of root causes for slow performance. How further quantitative analysis could explore a reciprocal or hyperbolic relation between total time to fetch records and pre fetch size and thereby optimum pre fetch size that can be recommended to resolve the performance issue.

### 8.1. About the case study application

### 8.1.1. Functional description

The bank acts as a custodian of assets deposited in the bank by their customers or account holders. The case study application facilitates bank's customers to view through web browser across internet and through bank's portal various reports of their assets like positions, balances, corporate actions of companies where bank's customers had invested in, their NAV, interest income etc.

### 8.1.2. Scenarios or use cases having low performance

The graphical user interface has many links for user to click and view reports with one link for a report. The response time to view majority of the reports in a browser was about 1.5 to 2 minutes and the bank's expectation or requirement is below 10 seconds response time. A response time of less than 3 to 6 seconds will be the expectations as a better user experience.

### 8.1.3. High level view of the system

Architecturally this is a 3 tier web application with the following technology stack.

- JavaScript, JSP, HTML, SmartClient for presentation layer

- Java application deployed in BEA WebLogic application server.

- Data base is Oracle and Oracle global data warehouse.

- Data access mechanism is through Java JDBC APIs

The deployment used and data flow is shown in figure 1 below.

*Fig 1: Existing deployment view and data flow*

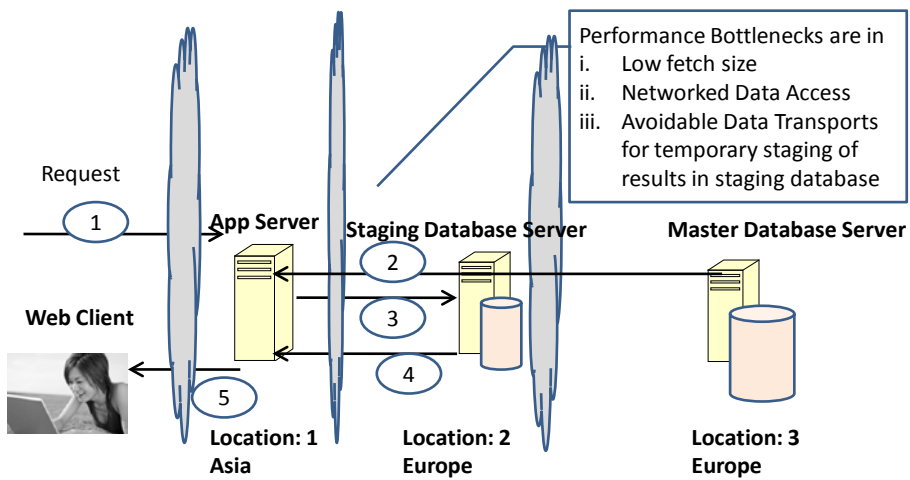

1 to 5 is the data flow sequence after each request

Data flow sequence to serve any user request for reports is depicted in figure 1. Application server fetches data records from global data centre in Europe (location 3) and first 500 records are temporarily parked in a staging data base in another different country in Europe (location 2). Using a generated SQL query as per user requests to further filter the data application server in Asia (location 1) fetches data records and presented as reports to the user browser.

### 8.2. Breakup of various stages in the end to end process flow in the system

From the time user login and enters requests for some report the flow is divided in to various stages as shown in figure 2. The flow depicted is common across all reports. Some of the use cases are

- User creates and stores an input template form, which can be used to input parameters for the report.

- User either uses a template created earlier or gives input in new form and retrieves a report.

*Fig 2: Blocks of processing stages*

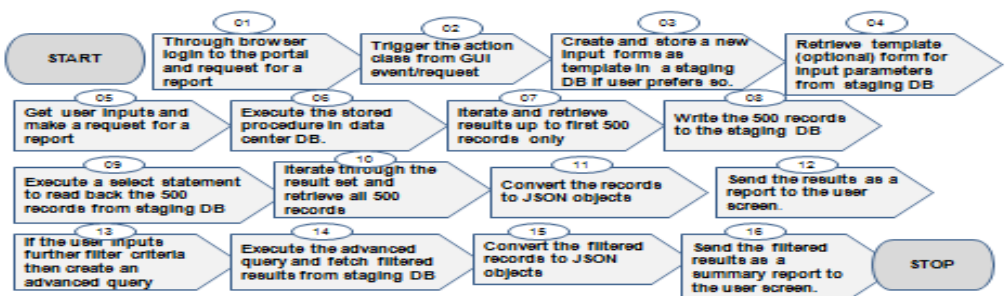

The elapsed time taken for each stage for various reports is measured by code instrumentations and profiling by manual code changes to log time differences across various processing stages. Most of the reports size is about 500 records or limited to 500 records while few summary reports are of size less than 10 records. The measured response time to view various reports of size 500+ records ranged from 1 minute to 2 minutes.

### 8.3. Inference from deployment architecture and time measured across various stages in process flow

From the measured elapsed time for each stage depicted in figure 2 and across various stages in process flow, we could observe that higher proportion of time was spent on data access from database in data centre and from staging database i.e. stage 6, 7, 9 and 10 in figure 2.

From the deployment architecture in figure 1, we could see that application server where the data was retrieved and processed was located at different geography from where the data base was. Network latency added delay in data access which could be reduced by about one half as measured in the case study by co locating application server and database.

Figure 3 gives further breakup of time between query execution and iteration through result set and retrieval of all records.

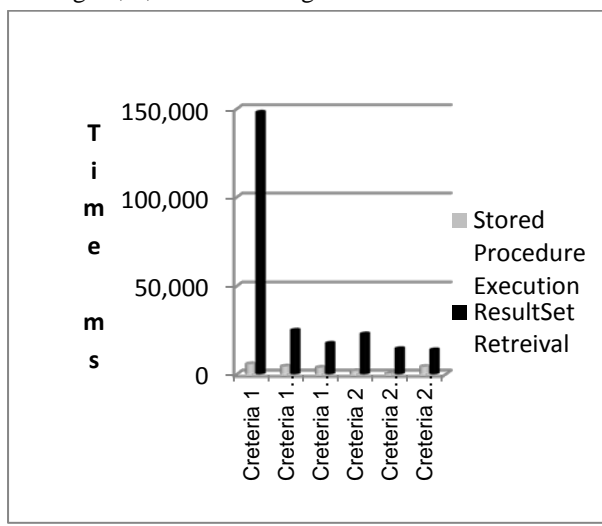
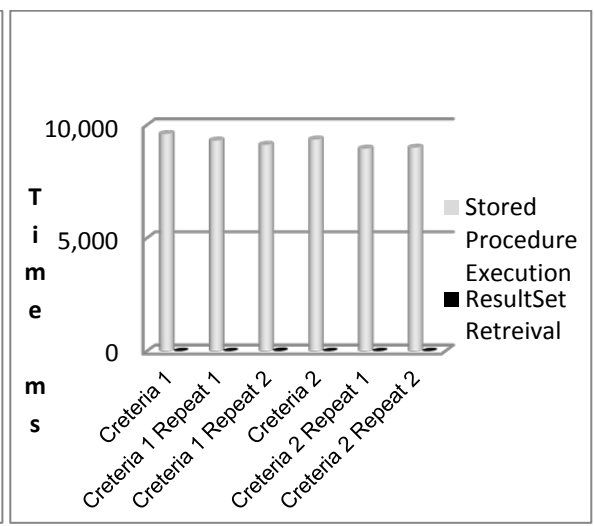

Fig: 3a: Stored procedure execution time and result set retrieval time. Result set has about 502 records.

Fig: 3b: Stored procedure execution time and result set retrieval time. Result set has about 5 records.

Figure 3 shows time measured in two stages as

1. Stage 1: Time to execute a callable statement and obtaining a [1]ResultSet, i.e. stored procedure execution time as measured to execute a Java statement resultSet = *cstatement.executeQuery() ;*

2. Stage 2: Time to iterate through the above ResultSet and extract all data records as java objects, i.e. result set retrieval time as measured to iterate through result set and retrieval of all data records as Java objects. i.e. the Java statement

*while ( resultSet.next( ) ) { //code to extract field objects from result set object }*

From the query execution time and results set retrieval time breakups in figure 3 we could see that, when number of records ( 502) are more (figure 3A), data record retrieval from result sets takes higher proportion of time compared to query execution time. When the number of records (5) is less (figure 3B), query execution time is of higher proportion compared to retrieval time. For the case study application majority of reports are about 500+ records as in figure 3A. Thus at entire application level data retrieval (transport)

IJCSI International Journal of Computer Science Issues, Vol. 8, Issue 6, No 3, November 2011
ISSN (Online): 1694-0814
www.IJCSI.org

401consumes higher proportion of time than query execution as well as any other operation or process.

Thus any effort on query optimization and database tuning for this application is not going to give any considerable performance benefit. Hence the focus of optimization in this application should be on optimizing on data transportation and reducing number of round trips between application and the database server rather than on the SQL query optimization.

Having narrowed down to data transport as a cause for delay, let us focus on overall data transportations scenarios in the case study application. From the process breakup (figure 2) and dataflow diagram (figure 1), we could see that for the purpose of doing further advanced filtering of data from already retrieved data records using SQL queries, data are again stored in a staging database. Based on user entered filter criteria a query is formed and filtered data is retrieved from staging database. This staging of data in another temporary database adds additional write to and read from staging data base adding to avoidable delays in data transport. This is a case of logical inefficiency because further advanced filtering can be done without staging the data in staging database.

## 8.4. Further drill down on data transportation to isolate root cause for higher data transportation time

### 8.4.1. Pre fetching rows from database to application server

In general a SQL query ResultSet object (a Java JDBC object in application layer) may point to 0 or 1 or thousands or even millions of records which needs to be pulled from database by iterating sequentially through the ResultSet using resultSetObject.next() call. When a SQL query is executed in a database through Java JDBC APIs running inside any application server, almost all database drivers also called as resource adopters have a feature to pre fetch more than just one database record to the ResultSet object of the application even if the request is to iterate and extract one single next record. Subsequent call or execution of the statement "resultSetObject.next()" will extract from pre fetched data records from resultSetObject and not from database server to avoid transportation delay across network between application server and the database server. Oracle has a default pre fetch size of 10 records and Sybase has 20 records.

### 8.4.2. Configuring pre fetch size

JDBC APIs like setFetchSize(int size) or configurations through application servers are only recommendations to the database driver and the driver may or may not enforce the recommendations. The JDBC APIs to get the pre fetch size i.e. getFetchSize() can give only the recommended value and not the effective value actually used by the driver.

### 8.4.3. How to identify effective pre fetch size

A simpler and practical way to know what is the effective pre fetch size is to measure the time taken to iterate through and retrieve each record and plot this series of time to fetch consecutive single record from ResultSet object against the sequence number of each consecutive record retrieved and look for peaks in the series. The peak occurs when all the pre fetched and cached records in result set object were already retrieved and to retrieve the requested record, the driver program has to fetch next set of records from database server and not from the local cache in result set object

Figure 4 shows the elapsed time to extract each consecutive record in the case study application, i.e. time to execute the statement **"resultSet.next()"** which is inside a while loop as in the code section below.

resultSet = cstatement.executeQuery() ;   //stored procedure execution time

while( resultSet.next() ) {     // record retrieval time plotted in Y axis and loop index in X axis

  //code to extract java objects from resultSet object

}

IJCSI
www.IJCSI.org

Fig 4: Retrieval time of consecutive records from result set

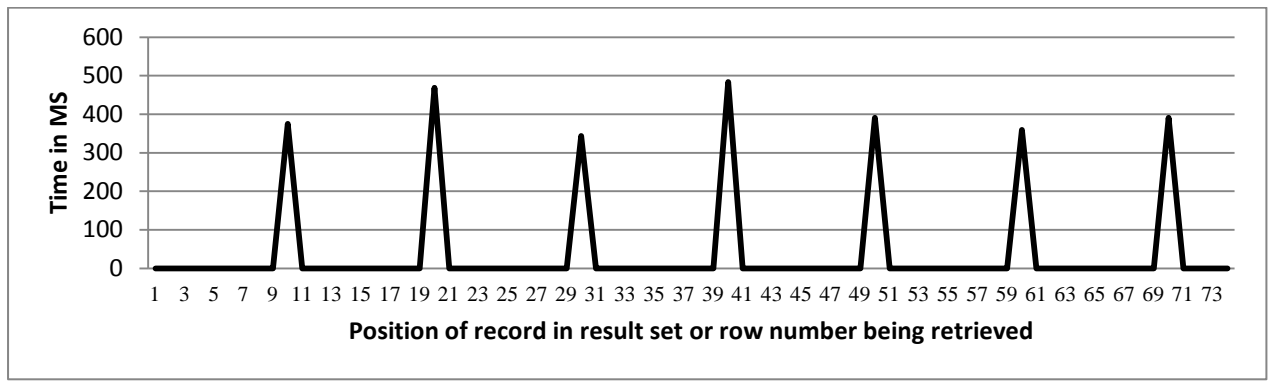

From figure 4, we can observe that, time to retrieve each consecutive record was nearly zero milli seconds and negligible, but suddenly increased and peaked to about 300 to 500 ms for every $11^{th}$ record. This implies that the JDBC driver pre fetched in batches of 10 records at a time and records 1 to 10 are retrieved from local resultSet object which is in memory retrieval and hence nearly zero milli seconds. However when the application attempted to iterate and extract beyond $10^{th}$ or $20^{th}$ or $30^{th}$ etc record, the driver again pre fetched next 10 records from database server which required one more round trip to database server from application server and had to transport a batch of 10 records of data even if the request was for next single record which explained the peak in time of about 300 to 500 ms to fetch $11^{th}$, $21^{st}$ and $31^{rd}$ records respectively. When we changed the pre fetch size to higher number by configuring pre fetch size in app server or through JDBC APIs, the effective pre fetch size was still 10. This is because setting pre fetch size through JDBC APIs or through app server deployment descriptors are just a recommendation to the driver and not guaranteed and has not changed the effective pre fetch size.

### 8.4.4. How to set and enforce desired pre fetch size

In the case study application when we used the Oracle JDBC extension class like OracleConnection, OracleResultSet and when we set the fetch size through Oracle extended APIs of these classes we were able to set and achieve desired pre fetch size. The effective pre fetch size can be experimentally observed by similar elapsed time versus row sequence number of records, wherein we could observe the peaks at every $21^{st}$ record when pre fetch size was set to 20 through Oracle JDBC extension APIs.

### 8.5. Root cause for low performance in the case study application

In the case study application the query execution time was less compared to records retrieval time. Since effective pre fetch size was the default 10 records and most of the reports were of 500 records, each report needed 50 round trips between database and application server which was identified to be the root cause for low performance.

Thus in the case study application for reports retrieval use cases many reports were of size 500+ records. From the processing stage breakup perspective, among many stages from user request to reports display data access stage was the main performance bottleneck. If we further break up data access in to 2 parts (I) time for query execution and (ii) time to retrieve data records, time to retrieve data records took more time. If we further break up data records retrieval, it was data transportation by multiple round trips between application server and database server which took higher proportion of time. Thus multiple round trips due to low pre fetch size were found to be the root cause for the slow performance in viewing reports.

### 8.6. Breakup of various stages of data retrieval to application server from database server using SQL query from within a Java JDBC application

Let us try to break up various internal processing steps from the time the application server issues query execute request through JDBC API till the query result set data base records of the executed query from database server are retrieved as collection of java objects in the application server or client.

From coding perspective using JDBC APIs, above data records retrieval involves two stages

1. Stage 1: Time to execute a statement or prepared statement or callable statement and obtaining a ResultSet, i.e. time to execute a Java statement

   resultSet = *cstatement.executeQuery() ;*

2. Stage 2: Time to iterate record by record through the above ResultSet object which is a pointer or handler to all the records of the query results and extract all data records as java objects, i.e. result set retrieval time from the code section below

   *while ( resultSet.next() ) { //code to extract field objects from result set object }*

However internally there may be many underlying processing steps and stages, which are explored below.

Assume that a java application deployed in an application server is making a JDBC call to execute a SQL query which has "N" records in the result set. N may vary from 0 to several millions and if N is very high because of the limitation of memory size in app server, the driver cannot bring and hold all the records in one shot in its result set object but may hold a cursor, a handler in the result set and may fetch a fraction of N records on demand. Thus the driver has to make multiple trips to the database driver when the application iterates through the ResultSet object and retrieves all the N records of the query. Since multiple trips across network is time consuming most of the database drivers optimize by pre fetching a finite number of records ahead even if the application iterates and request for just 1 record. Oracle has a default pre fetch size (f) of 10 records and it may vary with other databases like Sybase, DB2 and MSSQL server.

The breakup of various smaller stages in executing and retrieving "N" records to the application server from database server can be expressed as in Eq. 1 below. If we consider all the select statements to fulfill a use case in an application, then N may include records from multiple select SQL statements; however our focus here is the total read time of N records from a onetime single select statement execution and from single ResultSet Object.

$$T = \sum_{i=1}^{n}( r_i + e_i + a_j + t_i + c_i ) \quad (1)$$

Where

T = Total time to retrieve all N records of the result set to the application server as java objects from the database server.

The subscript "i" can change from 1 to n, where n is the number of round trips the driver had made to bring all the records to the ResultSet object pre fetching 'f' records in each trip. If 'N' is the total number of result set records for the select query and 'f' the effective pre fetch size, then n = N/f + (1 if (N modulus f > 0)). For example if the query result set has N = 502 records and the effective pre fetch size (f=10) then the driver may make 51 (50 to bring 500 and 1 more trip to bring the residual 2 records).

$r_i$ is the elapsed time to make a request from app server to the database server during the $i_{th}$ trip.

$e_i$ is the query execution time spent in the SQL engine running in database server.

e = Hard parse time + soft parse time + search time to select each record meeting the select query criteria + traversal time to locate the searched record + seek time to fetch and join records from multiple tables + time to read records from disk store to the memory of SQL engine.

The SQL engine may do a onetime execution, when i = 1 and may retrieve and cache the records in SQL engine cache also called as database server side caching. Subsequent retrieval of records may be from the SQL engine cache in the database server. This execution time includes hard parse time and a soft parse time. [3]Parse time includes loading SQL statement to memory, a onetime and first time syntax verification of the SQL statement, authorization to access the tables, creating and optimizing execution plan and actual execution time of the select call. Hard parse is relatively more expensive in terms of CPU time. Execution time e will increase if more number of tables are joined, large numbers of records are joined, search on non indexed fields etc.

$a_j$ is the access time spent during the j th iteration by the SQL engine in the database server to retrieve a fraction or all of the N records to SQL engine cache from the database server disk store. SQL engine caching size j referred to as server side cache may be different from the pre fetch size f which is client (app server) side cache. Maximum number of iterations to retrieve all N records from database store to database engine, i.e. upper limit of j depends up on SQL engine's caching capacity in database server whereas the upper limit of i depend upon on the JDBC driver's effective pre fetch size (f) and application server memory availability.

$t_i$ is the elapsed time per trip to transport f records from database server to the application server in i$^{th}$ trip.

$$t_i = f1(\beta_i, h_i, b_i, a_i) \quad (2)$$

Where transport time $t_i$ on ith trip is a function f1 of $\beta_i$ network bandwidth, which will be of the order of few megabytes per second. This transport time decreases with increasing β

$h_i$ number of network hops in ith trip. The application server and database server in data center may sometime be in different location and may even be in different country and hence data has to be transported by more than one network hops. It should be noted that it is not the distance between app server and the database server, but the number of network hops which is relatively significant and important factor in transportation time.

$b_i$ is the number of bytes of data being transported during ith trip. This is the sum of all individual field size in bytes in a record multiplied by the by pre fetch size f (the number of records transported in i$^{th}$ trip.

$a_i$ on wire network bandwidth available due to bandwidth being shared by many different application or process concurrently.

Data transportation time can be expressed as

$$t_i = \sum_{h=0}^{h=k} t_i^{k(\beta)} \quad (3)$$

Where, k is the total number of network hops the system has made to transport the pre fetched f records during i$^{th}$ trip between application server and database server. Obviously k = 0, when both application server and database server are same, which is generally not the case in real production level systems. Dependency β in the equation (3) implies that the network bandwidth may vary between each network hop and hence the transportation time may vary in each hop.

$c_i$ is the elapsed time to convert from data base specific data types to java data typed objects. This includes conversion of data in each cell or field of each record and for all the f records, the application fetched in i$^{th}$ trip. For result set in a single select query, since the record structure (number of columns or fields, data type in each column) is same for all the f records, the subscript i in $c_i$ can be removed. In the equation subscript i is retained because the record structure may vary across different select queries in the application. The parameter c may include marshaling data base objects on wire and un-marshaling database specific objects to Java objects by the database driver. Some driver may cause unusually large delays or even error if incompatible data types between database and Java are converted.

In Eq. 1 the factors $e_i$ query execution time and $t_i$ transportation time are two major time consuming factors and in any application either one or both of them may be the dominant time consuming factor. The optimization on either query or transportation or both can be focused accordingly.

Since in our case study application, data transportation time is of higher proportion and found to be the root cause as described in section 8.5, let us focus and elaborate on data transportation components and terms of Eq. 1

## 8.7. Relation between Elapsed Time and Pre Fetch Size

If we split the data access in application JDBC layer in to two parts as described in section 8.4.4

- stage 1: statement execution time and

- stage 2: data extraction time from query result set which includes predominantly data transportation time to transport data records in batches of f records being pre fetched during each trip and isolate and measure only the data transportation time, then

$$T_2 = \sum_{i=1}^{r} t_i \; \alpha \; R$$
(4) and

$$T_2 = \sum_{i=1}^{r} t_i \; \alpha \; f$$
(5) where LHS is the total transportation time and R is the number of round trips between application server and database to fetch all records of the query result set. In other words total retrieval time is proportional to number of round trips Eq. 4 and also proportional to the size f of data being transported in each trip Eq. 5.

Combining equation (4) & (5), we have

$T_2 = K_1 R + K_2 f$
(6) Since R = N/f if N is integral multiples of f, otherwise R = N/f + 1 where, N is the total number of records for a given query. Rewriting Eq. 6, we have

$T_2 = K_1 N/f + K_2 f + K_3 + K_4$ (N modulus f)
(7)

Where

$K_1$ is the proportionality constant of time variation with number of roundtrips when f is the number of records being fetched in each round trip between database and app server. $K_1$ is average time spent per trip when the data size is of f records.

$K_2$ is the proportionality constant of time variation with data size. $K_2$ is the average time to transport data size of f records.

$K_3$ is the average time per trip, when data size is of (N modulus f) records

$K_4$ is the average time to transport data of size equivalent to (N modulus f) records

First and third term of Eq. 7 are time spent due to N/f and 1 round trips respectively to transport data records and $K_1$ and $K_3$ are proportionality constants of time per trip when size is f and (N modulus f) respectively. Similarly second and fourth terms of Eq. 7 are time spent on transportation due to data size and $K_2$ and $K_4$ are proportionality constants of time spent to transport data of size 1 record in single trip. K2 and K4 can be considered as almost equal.

$K_1$, $K_2$, $K_3$ and $K_4$ can be determined by regression by collecting elapsed time data for various values of number of round trips and data size.

First term in Eq. 7 is rectangular hyperbolic or reciprocal variation of time with pre fetch size, while second and fourth terms are near linear with f. The curve between elapsed time T2 and the pre fetch size f will be either rectangular hyperbolic or linear depending upon which of the two component viz. number of round trips or data size is dominant in Eq. 7. Let us find out the dominant component and term in Eq. 7 from general observations as well as from the case study application measurements in next section.

### 8.7.1. Comparison of impact of record size and number of round trips on performance

Table 2 gives the data transportation time measured in the case study application for different values of number of round trips and different values of pre fetch size.

*Table 2: Impact of data size and number of round trips on data transportation performance*

|  | When pre fetch size f = 10 records | When pre fetch size f = 252 records |
|---|---|---|
| **Impact of data size on performance:** Average time per trip to retrieve f records in one trip between app server and database server | 450 milli seconds (per 1 trip) | 650 milli seconds (per 1 trip) |
| **Impact of number of round trips on performance:** Total time to retrieve all 502 records in multiple round trips with pre fetching f records per trip between app server and database server. | 14 seconds (in 51 round trips) | 2 seconds (in 2 round trips) |

From table 2 we could see that, the reduction in time due to reduction in number of round trips between app server and database server is very high (from 14 seconds to 2 seconds) compared to minor increase of time of few milli seconds (from 450 milli seconds to 650 milli second) due to increase in data size. Thus the total time to retrieve all records in a result set is pre dominantly determined by number of round trips expressed in Eq. 4. Since round trips R = N/f where N is the total number of records in a result set and f the pre fetch size after ignoring the data size factor and also the time to transport the residual last 1 trip with N modulus f records. I.e. ignoring all terms in Eq. 7 except the first term. We can write retrieval time T2 as a function f2 of number of round trips neglecting the effect of data size expressed in Eq. 5 and 7, i.e.

$$T_2 = f2(R) = f2(N/f)$$

$$T_2 \, \alpha \, N/f$$

$$T_2 = K_1 \, N/f \quad (8)$$

$$K_1 = f3 (\beta, h) \quad (9)$$

Where $K_1$ is the proportionality constant of time per round trip when the data size is of f records in each trip. $K_1$ is assumed a function f3 of network bandwidth β and number of network hops h. Eq. 8 obtained after approximations described above on Eq. 7 is the reciprocal or rectangular hyperbola relation between T2 and f.

### 8.7.2. Impact of network bandwidth and network hops on performance

Table 3 gives the data transportation time measured between two different network paths from data base to i) app server deployed in Asia and ii) app server deployed in Europe.

*Table 3: Impact of network bandwidth and network hops on data transportation performance*

| Effective Pre fetch size | Time to transport 502 records of query Q with Data centre in Europe and Application Server in Asia | Time to transport 502 records of query Q with Data centre in a country in Europe and Application Server in another country in Europe |
|---|---|---|
| 10 records | 14 seconds | 7 seconds |
| 50 records | 8 seconds | 4 seconds |

From table 3 we could see that when all other parameters remains same, time to transport across networks between Europe to Asia is twice that of time to transport records across networks between one European country and another European country. Only difference between these two cases is number of network hops and network bandwidth. We could observe almost a parallel curve of nearly half the time for data transport between the two European countries as compared to transport time between the European country and the Asian country (figure 5).

### 8.7.3. Rectangular hyperbolic decrease of total elapsed time to transport a set of records with increasing pre-fetch size

Eq. 8 implies a reciprocal or rectangular hyperbolic relation between $T_2$ total time to transport all (N) records of a result set and the pre fetch size f. We can see from Eq. 8, that when f tends to zero, time tends to infinity and when f tends to infinity time tends to zero. Thus asymptotes are parallel to Y (Time $T_2$) and X (fetch size f) axis.

Figure 5a is a theoretical curve of Eq. 8 with assumed value of $K_1 = 0.2$. Pre fetch size f is plotted in x-axis and (0.2 * N/f) * 1000 in y-axis. N is taken as 502 and 1000 is multiplied to show time in micro seconds. In the figure 5a only positive values of 'f' is plotted as practically negative f has no meaning.

Figure 5b is the experimental curve measured from the case study application and shows how the total elapsed time "$T_2$" plotted in vertical y-axis to fetch a report containing about 502 records from database to application server reduces with increasing pre fetch size "f" plotted in horizontal x-axis.

We can qualitatively see the non linear rectangular hyperbolic shape and trend of $T_2$ vs. f curve measured from the case study application. When the pre fetch size is 0, then the application can never fetch any data and hence takes infinite amount of time and the curve will be parallel to (T2) y-axis. Similarly when the pre fetch size f is relatively high compared to N, for example 200 in our case study application, application has to make 3 round trips to fetch 502 records and if pre fetch size is increased by 1 unit, the application again needs only 3 round trips to fetch all 502 records. I.e. decrease in number of round trips is 0 per unit increase in pre fetch size when f = 200 or in other words slope is nearly zero and parallel to x- axis at f = 200. In other words the magnitude of slope of the curve is very high or the slope tends to very high negative or minus infinity when f tends to 0 and the magnitude of slope rapidly but smoothly reduces to zero, when f tends to N.

Thus the experimental curve 5b has similar properties of theoretical curve 5a of rectangular hyperbola and the Eq. 8.

Another property of $T_2$ vs. f reciprocal curve is the decreasing slope with increasing f. The slope important and worth to study because

1. Negative slope indicates that elapsed time $T_2$ to fetch records decreases with increasing f.

2. Decreasing magnitude of slope with increasing f indicates that the rate at which elapsed time $T_2$ decreases or the gain in performance for a unit increase in fetch size f is relatively high when f is small and the gain in performance for a unit increase in f is relatively small when f is large.

We can see how the slope decreases with increasing f by comparing the slopes at different points on the curve in figure 5b which is detailed from case study measurements in the next section.

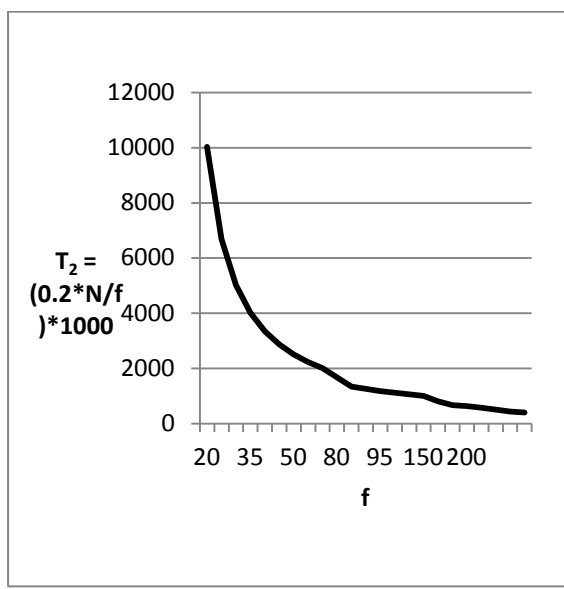

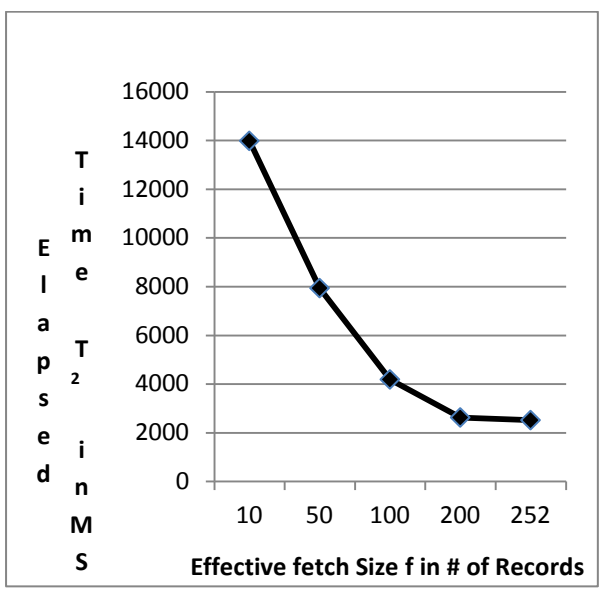

*Fig 5a: Theoretical curve of $T_2 = (0.2*N/f)*1000$ Vs f*

*Fig 5b: Experimental curve from case study showing decrease in elapsed time ($T_2$) with effective fetch size (f)*

## 8.7.4. Understanding the reason for decreasing magnitude of slope with increasing pre fetch size of transportation time vs pre fetch size curve

Decreasing slope with increasing f in $T_2$ vs. f curve (fig 5b) for a given N can be understood by calculating and focusing on the decrease in number of round trips achievable per unit decrease in f at different points in the curve for increasing values of f from low value of f to high value of f. From simple illustrative calculation shown in table 4 for a given N = 502, we can see that when f is low there is a high reduction in number of round trips and hence higher gain in performance even for a small or unit increase in f but when f is higher less reduction in number of round trips and hence less significant performance gain by same quantitative or unit increase in f.

*Table 4: Illustration of decreasing slope with increasing pre fetch size*

| Seq # | $f_1$- Effective Pre-fetch size in number of records | $n_1$ -Number of round trips to fetch 502 records when pre fetch size is $f_1$ | $f_2$- Effective pre fetch size increased by 1 | $n_2$ -Number of round trips to fetch 502 records when pre fetch size is $f_2$ | $(n_1-n_2)$ - Decrease in number of round trips to fetch 502 records per unit increase in f at $f_1$. | Slope of $T_2$ vs f curve. Decrease in time in ms per increase in f by 1 record. Assuming average of 400ms per round trip. |
|---|---|---|---|---|---|---|
| 1 | 0 | Infinity | 1 | 502 | Infinity | Infinity |
| 2 | 1 | 502 | 2 | 251 | 251 | 100400 |
| 3 | 10 | 51 | 11 | 46 | 5 | 2000 |
| 4 | 100 | 6 | 101 | 5 | 1 | 400 |
| 5 | 168 | 3 | 169 | 3 | 0 | 0 |
| 6 | 200 | 3 | 201 | 3 | 0 | 0 |

Column 6 in table 4 shows how the reduction in number of round trips per unit increase in f at various values of f to fetch 502 records decreases with increasing f from infinity to zero rapidly as f is increased from zero to 168. One can see from table 4 that for a total number of records of 502, when f is low like 1 record, even for a small increase of f to 2 records, the reduction in number of round trips to fetch all 502 records reduces from 502 to 251. One can compare this with reduction in number of roundtrips by only 5 for a same unit increase in f, when f is 10 and a reduction of just 1 round trip by unit increase in f when f is 100. Thus the performance gain per unit increase in pre fetch size is very high when f is low and the performance gain reduces rapidly as f increases. Performance gain is very low per unit increase in f when f is higher and reaches zero after f >= 168 for N = 502.

To generalize for any values of total number of records to retrieve N, one can say that, when the ratio (N/f) between number of records N and pre fetch size f is high, even a small change in f will bring large benefits in performance. When the ratio N/f is small, even large change in f will not get considerable performance benefit. This is in consistent with the reciprocal nature of the Eq. (8) and the theoretical curve of figure 5a.

Figure 6 shows the plot of the slope $dT_2/df$ in y-axis and f in x-axis, where $T_2$ is the total elapsed time to transport about 502 records observed in the case study application and f is the effective pre fetch size in number of records.

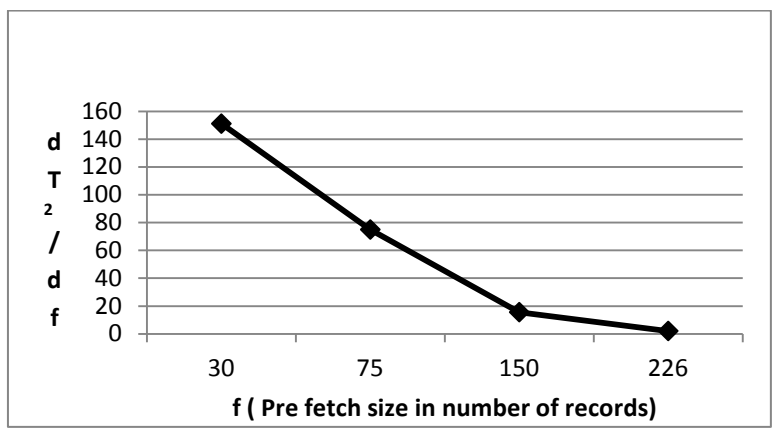

*Figure 6: dT$_2$/df (Slope of Elapsed Time With Pre Fetch Size)*

Figure 6 is the derivative of curve of figure 5b, where we can observe that the rate of decrease of elapsed time with pre fetch size (slope) decreases with increasing pre fetch size.

### 8.7.5. Differences between theoretical rectangular hyperbolic curve and actual transportation time vs. fetch size curve

Though there are many similarities between theoretical rectangular hyperbolic curve as in Eq. or figure 5a and actual $T_2$ Vs. f curve, there are many differences. Few differences are listed below.

Eq. 8 is only an approximation after removing size factors and other terms in Eq. 7, hence actual $T_2$ vs f curve will have slight lift towards higher $T_2$ which increases with f at each point in the curve when compared with theoretical curve (ref fig. 5a and fig. 5b). The lift is due to size factor of slight increase in time to transport higher data size due to higher number of records (f).

For theoretical rectangular hyperbolic curve, the 'f' in Eq. 8 has to be continuous, whereas practically f is discrete.

Also at certain values of f for a given N, $T_2$ may not decrease even when f is increased unless increase in f results in decrease in number of round trips. For example assume that N = 502, and f is increased from 251 to 252. Number of round trips is same and equal to 2 for both values of f. Comparing time $T_2$ when f = 251 and f = 252, one can see that, time to transport 251 records twice may be of same value or may not decrease when compared with time to transport 252 records in first trip and remaining 250 records in 2$^{nd}$ trip.

Thus the reciprocal or rectangular hyperbolic nature of $T_2$ vs. f is only a dominant trend and an approximation and not an absolute relationship between $T_2$ and f. However rectangular hyperbolic trend between transportation time and pre fetch size can be treated as a generic dominant and approximate trend for any application accessing relational data through JDBC and will help in performance tuning and deciding optimal pre fetch size. If N is known and K1 is determined, then this relation can give an approximate estimate of quantifiable expected performance gain achievable for various values of f without much trial and error.

### 8.8. Trade off between higher pre fetch size and memory consumption

Though performance gain can be achieved by reducing the number of round trips between application server and database servers by increasing the pre fetch size, higher pre fetch size requires higher memory allocation in application server. In the case study application, when the pre fetch size was increased from default 10 records to 500 records as most of the report size was about 500 records, the application server was found to crash with out of memory error exception. Thus there is a tradeoff between performance gain by higher pre fetch size and higher memory consumption.

### 8.9. Threshold pre fetch size

From the figure 5 and 6 as well as from the rectangular hyperbolic nature of $T_2$ Vs f curve, we can see that, the curve (very high negative slope) is almost parallel to Y axis ($T_2$ axis) when X (f) is near zero. Thus there is a rapid decrease in elapsed time ($T_2$) with increase in pre fetch size (f) when pre fetch size was low. However beyond certain f, the curve (has very low or near zero slope) is almost parallel to X axis (f axis) and hence very less or near zero decrease in elapsed time ($T_2$) per unit increase in f when f is large. For the case study application we can see that from figure 5b, the curve became parallel to x axis (f axis) for values of f above 200 and from figure 6 we can see that the slope (rate of decrease of elapsed time with pre fetch size) already became zero when f = 226. Thus there going to be no significant performance gain by increasing pre fetch beyond 226 records. Since pre fetch size of 226 requires 3 round trips (226 + 226 + 50) to fetch all 502 records, with 502/3 = 167.3, i.e 168 (168 + 168 + 166) fetch size, we can fetch all 502 records in same 3 round trips. The performance loss due to slight increase in time required to fetch when the data size increases due to higher fetch size being less significant compared to performance loss due to increased number of round trips as round trip is being considered as a main tuning parameter. With same number of round trip lesser pre fetch size will have lesser memory footprint in app server hence 168 should be the preferred pre fetch size than 226. Pre fetch size of 168 can be considered as the threshold or optimal pre fetch size for this case study application where the number of required records N is limited to 502.

## 9. Recommendations for tuning the case study application

Based on elicitation and extraction of application architecture, deployment architecture, data and process flow logic and analysis of the same, time measurements by code instrumentation, profiling of various major stages of execution, identification of causes and solutions and relative benefits of solutions the following recommendations were made.

*Table 5: Proposed performance tuning recommendations for the case study application*

| Identified performance bottleneck | Suggested solution as changes | Level of abstraction of the change and estimated effort in person days | Trade off or factors to consider before making implementation decision | Estimated and measured benefits in trial implementation of recommended solution/changes. |
|---|---|---|---|---|
| Application layer is making avoidable multiple (50) round trips between app server and data base as the pre fetch size was default 10. | Increase the pre fetch size to the threshold value of 168. Use Oracle Connection, Oracle Statement and Oracle Result Set to effectively set the pre fetch size as setting pre fetch size in application server or setting pre fetch size through JDBC Connection, Statement and Result set were ineffective. | Code level. 2 | Higher pre fetch size demands higher heap memory in application. Using Oracle specific extension APIs Vs generic APIs. | Measured benefit from average 70 seconds to 25 seconds per report retrieval. |
| Application server and data center are at geographically different locations leading to higher network latency to access data due to more number of network hops and | Move application server to the location of data center to co-locate data and application consuming the data. Moving database to application location is expensive and hence the | Deployment Architecture. 3 | Moving application server from existing Asia location to Europe location where data center is implies relocating application IT team to Europe or | From 25 seconds to 12 seconds per report retrieval as time was found to reduce by about one half for retrieving many reports. |
| hence delays. | suggestion of moving application server to data center. | | doing remote support. | |
| There was an extra write and read operation due to staging result sets in a staging database | Change the application logic to directly read the data from main data base and avoid staging database. | Logic and Design. 5 | With retrieved data temporarily staged in database further filtering on retrieved records can be easily done by SQL query filters. Avoiding staging database implies doing data filter in Java. | This extra write and read operation was estimated to consume 4 seconds per report retrieval. Estimated time gain is from 12 seconds to 8 seconds. |
| User views only first 500 records, whereas the stored procedure extracts all and more than 500 records meeting the query criteria. Some queries had even 40 thousand records. | Change the stored procedure to limit the number of records to first 500 among all records meeting the query criteria. | Code/SQL script: 3 | Unused records are being fetched in database wasting time and memory. This change has no trade off. | Not estimated. |
| Field (column) size of record in the table is not limited to required size like 50 characters length in Java. Instead it uses table field default size of 4000 characters per VarChar2 field. This causes huge memory allocation when records are fetched into app server from database. | In Table definition use varchar2(50) instead of varchar2. This indicates to the driver to allocate only 50 characters instead of 4000 characters per column of type Varchar2. | Data Model/Table Structure. 2 | More than necessary field width is consuming avoidable memory. | This reduces the memory demand in app server while reserving memory to store result set records and enable us to use higher pre fetch size. |

## 10. Conclusion

Re engineering an existing application in production is different from engineering for developing a new application for performance or for any other quality of service. There is a growing need to re engineer many business applications already deployed and serving in production for higher performance due to increasing demands. Performance problems can be in any of several processing or data flow stages like user inputs, network, processing, data access etc. Solutions can be applied as changes at architecture, design, framework, logic and coding level. An inquisitive and experimental approach starting with higher abstraction (e.g. architecture) level view to identify causes and solutions, then a drilled down (e.g. design) level causes and solution and further drill down to identify root causes and respective solutions will help. More than one factor can cause performance degradations hence estimating relative benefits will be helpful. Cost benefit analysis requires measured or sampled or estimated quantitative benefits and trade offs for each of the solutions. It is common for multi tired business applications with data access as the common performance bottlenecks area. Though there can be a slight increase in transportation time with data size, there is a dominant reciprocal or rectangular hyperbolic relationship between total transportation time to retrieve result set records of a query and the pre fetch size, the number of records the database driver brings to the client and caches at client side from database server. Data records retrieval time reduces more rapidly with increasing pre fetch size when the ratio of number of records to retrieve to pre fetch size is high however the performance gain is negligible when this ratio is low. There is a threshold value of pre fetch size, where transportation time Vs pre fetch size curve appears to approach a point of inflexion where slope tends to zero when pre fetch size is increased further. Beyond this point, increasing pre fetch size will not bring any considerable performance benefit.

## 11. Acknowledgments

Author thanks Anupam Sengupta of Tata Consultancy Services Ltd for his effort in reviewing this paper and his valuable and pointed technical comments has helped me to make many corrections and improvements from initial version. Author thanks Banisha M, Syed Irfan Pasha and Saiaparna Kunala of HCL technologies who were part of case study application development team and helped me in analyzing the case study application. Banisha explained the implemented architecture and design of the case study application. Syed developed modules to measure time to iterate through and retrieve each record from ResultSet to identify the effective pre fetch size and Saiaparna instrumented the application to measure breakups between query execution time, query results retrieval time and to profile time across various stages of processing which helped us to isolate the performance hot spots.